\documentclass[11pt]{article}
\usepackage[margin=1in]{geometry}
\usepackage{authblk}
\usepackage{url}
\usepackage{indentfirst}
\usepackage[allcolors=blue]{hyperref}
\usepackage{xcolor}
\usepackage{booktabs}
\usepackage{threeparttable}
\usepackage{graphicx}
\usepackage[numbers]{natbib}
\setcitestyle{numbers,sort&compress}
\begin{document}

\title{Evaluating Phylogenetic Comparative Methods\\ under Reticulate Evolutionary Scenarios}
\author{
Lydia Morley$^{1,*}$, Emma Lehmberg$^{1,2,3}$, and Sungsik Kong$^{4,5,*}$\\
{\small $^{1}$Department of Ecology and Conservation Biology, Texas A\&M University, College Station, TX, 77801}\\
{\small $^{2}$Department of Biology, Lakehead University, Thunder Bay, ON, P7B 5E1}\\
{\small $^{3}$IISD Experimental Lakes Area, 111 Lombard Avenue, Suite 325, Winnipeg, MB R3B 0T4, Canada}\\
{\small $^{4}$The Institute for Computational and Experimental Research in Mathematics, Brown University, Providence, RI, 02903, USA}\\
{\small $^{5}$RIKEN iTHEMS, Wako, Saitama 351-0198, Japan}\\
{\small $^{*}$Co-correspondence to be sent to: Lydia Morley, 534 John Kimbrough boulevard, College Station, TX, 77843, USA, E-mail: lydia.morley@tamu.edu; and Sungsik Kong, 2-1 Hirosawa, 2nd Floor, Wako, Saitama 351-0198, Japan, E-mail: sungsik.kong@gmail.com}
}
\date{}

\maketitle
\vspace{0cm}
\begin{abstract}
{Phylogenetic comparative methods (PCMs) are widely used to study trait evolution. However, many evolutionary histories involve reticulate evolutionary scenarios, such as hybridization, that violate core assumptions of these methods. In this study, we evaluate how such violations affect the performance of PCMs. In particular, we focus on the ancestral character estimation, evolutionary rate estimation, and model selection. 
We simulate continuous trait evolution on various phylogenetic network topologies and assess the performance of PCMs that assume a bifurcating tree (i.e., major tree of the network) as the underlying model of evolution. We found that the performance of the tested PCMs was suboptimal. Using random forest, generalized linear models, and model-based clustering, we identified key factors contributing to these inaccuracies. Our results show that frequent and/or recent hybridization accompanied by one ore more transgressive events and rapidly evolving traits (i.e., high evolutionary rate) lead to significant estimation error, especially with respect to rate estimation and model choice. These factors substantially shift trait values away from tree-based model expectations, leading to overall increased error in parameter estimates. 
Our study demonstrates cases in which PCMs that rely on trees are likely to misinterpret biological histories and offers recommendations for researchers studying systems with complex evolutionary histories.}\\

\end{abstract}

\noindent Phylogenetic comparative methods (PCMs) are a set of inferential techniques designed with a common goal: discovering biologically (and causally) meaningful relationships between observed traits and evolutionary patterns (i.e., phylogenies) \citep{OMeara2012}. Though PCMs are traditionally \emph{tree-based}, as they are applied to phylogenetic trees, their application is expanding to phylogenetic networks \citep{Bastide2018}.
While these \emph{network-based} PCMs have shown great promise in improving estimation accuracy when traits are inherited through reticulate evolution (see \cite{Hibbins2021Introgression} and \cite{Wang2021Phylogenomic} for instance), they are still in their infancy and conventional tree-based approaches continue to dominate in practice \citep[][]{deWittSmith2008,Cardillo2015,King2015,Harris2022,Folk2023}.
Given PCMs' popularity and broad applications, it is important to understand whether and why interpretations of tree-based PCMs may be misled by reticulate evolution, especially as growing evidence highlights the prevalence of reticulate histories across the tree of life \citep[][]{Anderson1949,Ellstrand2000,Hovick2014,Toews2016,Leighton2021}.

Phylogenetic comparative methods emerged when \cite{Felsenstein1985} identified a fundamental issue in comparative datasets: many observed traits appear correlated simply due to phylogenetic inertia. For example, traits ${i}$ and ${j}$ might only be correlated because they independently evolved in a single common ancestor of the lineages that possess them, not because the evolution of trait ${i}$ `caused' the evolution of trait ${j}$, or vice versa. 
To address such statistical non-independence in phylogenetic datasets, \cite{Felsenstein1985} introduced phylogenetically independent contrasts (PICs). This method accounts for shared ancestry when testing trait correlations, paving the way for a phylogenetically explicit understanding of trait interactions, ecologies, and organismal biology. 
Today, a wide selection of novel, innovative PCMs exist, many of which are designed not only to calculate correlations between traits, but also to assess phylogenetic signal (i.e., the extent to which relatedness explains variation in a trait; \cite{Martins1997}), discover patterns and rates of trait evolution \citep{Hibbins2023Phylogenomic} including species phenotypes, ranges, and niches \citep{OMeara2006}, and/or rates of speciation, extinction, and diversification \citep{FitzJohn2012}.

The principles of ancestral character estimation (ACE; the estimation of phenotypic trait values for ancestral nodes in a phylogeny) are central to PCMs, such as estimating associations between character transitions and diversification rates \citep[e.g., Hidden State Speciation and Extinction (HiSSE);][]{Beaulieu2016hisse}, predicting historical areas of geographic occupancy (e.g., BioGeoBears; \cite{Matzke2013biogeobears}), estimating phylogenetic signal (e.g., Pagel's Lambda \citep{Pagel1999lambda} or Bloomberg's $K$ \citep{Blomberg2003}), and analyzing correlated evolution among focal traits with PICs \citep{Felsenstein1985}. Methods relying on ACE principles often strictly assume that each branch on a phylogenetic tree evolves independently, and according to a specified model (but see \cite{Nuismer2014Predicting,Drury2016CompetitionTraitEvolution} and \cite{Manceau2017Unifying} for PCMs try to accommodate for non-independence on trees). In other words, most tree-based PCMs assume that trait states on one branch do not impact trait states on another after divergence. Traits generated under a phylogenetic network violate this assumption, leading to branch non-independence when previously diverged edges unite at a reticulation node. Such a violation is expected to lead to biased ACE, as well as model and parameter estimates. 


An integral part of the ACE pipeline involves defining a model of trait evolution. The preferred model is typically chosen based on Akaike Information Criterion (AIC) or Bayesian Information Criterion (BIC), and that choice often informs the inference of some mechanistic relationship between a trait and its evolutionary history. Two widely used models are Brownian Motion (BM; \cite{Harmon2019}) and Ornstein-Uhlenbeck (OU; \cite{butler2004}). 
Brownian Motion models trait evolution as a ``random walk", in which values may change in any direction or magnitude at a given point in time. Brownian Motion is one of the simplest models of trait evolution and patterns of change are entirely described by two parameters: the trait value at the root of the tree (or starting value) and variance/rate ($\sigma^2$). In effect, BM describes a process in which traits evolve randomly, constrained only by phylogenetic inertia.
On the other hand, Ornstein-Uhlenbeck generalizes BM, where traits still evolve randomly, but they are more likely to evolve towards a particular optimum, with an additional parameter that determines the strength of the pull towards that optimum. Consequently, if model selection favors OU over BM, we may expect directional or stabilizing selection as a possible biological process that influenced the evolution of a focal trait. However, if the assumption of branch independence is violated due to processes like reticulation, BM may be falsely disfavored simply due to the hidden structure of the true phylogeny. 
In other words, tree-based PCMs may be more likely to select incorrect evolutionary models when applied to groups with suspected reticulate histories, making it difficult to draw biological interpretations based solely on model-fit statistics.
Comparable biases in model selection in trees, whereby simpler models are disfavored relative to more parameter-rich models under certain conditions (e.g., \cite{Cooper2015OUCaution,Silvestro2015MeasurementError}), further complicates this problem.
Moreover, the rate parameter estimated in BM can also be interpreted as biologically significant, meaning that not only model selection, but also model parameter estimates, may influence conclusions about ecological or intrinsic selective factors driving observed variation. Critically, without knowing how reticulation impacts model selection and parameter estimation, the reliability of such conclusions should be scrutinized. 
 
Aside from the fact that reticulations in a network violate the assumption of branch independence, variation in structural characteristics within a network may also contribute to differences in the magnitude of error when relying on tree-based PCMs.
One reason is that non-treelike evolutionary histories can manifest in multiple ways, some of which may be more likely to generate misleading signals. Three topological characteristics that are likely to influence ACE estimates are: (1)\textit{ the type of reticulation}. If reticulate evolution results in hybrid speciation, then estimating ancestral traits and rates of evolution might be best done under network models. On the other hand, if reticulate evolution results in repeated backcrossing into one parental lineage without the generation of a novel hybrid species (i.e., introgression), the proportional parental contribution to introgressed lineage might determine how severely traits are displaced following hybridization. (2) \textit{The age of the reticulation event}. Assuming most gene flow events are relatively recent, using tree-based models might result in erroneous estimations of characters at more terminal nodes, as well as evolutionary rate overestimation.
Errors occurring at terminal nodes may propagate toward the root, adversely affecting the accuracy of estimates throughout the tree.
(3)\textit{ The abundance of reticulation.} As number of reticulation increases in the network, we should expect to see decreased accuracy regarding evolutionary rates, ACEs, and best-fit models of trait evolution. Depending on the evolutionary history containing one or more of the above listed characteristics, the magnitude of error in the estimates may be more severe in some scenarios than others.

In addition to structural variations in a network, the evolutionary history of the focal trait itself may impact inference accuracy using tree-based PCMs. For example, rapidly evolving traits may allow for signals of reticulation to disappear faster via repeated convergence, or they may displace characters further, leading to increased error. Moreover, transgressive evolution alone can generate misleading signal. When a hybrid offspring possesses a phenotype that is outside of the trait-space of either parent, tree-based PCMs might be prone to overestimate base evolutionary rates, incorrect ACEs, and support a model that accommodates shifts in trait space outside of expectations under BM. 

In reality,  topological structure and trait history jointly interact, potentially amplifying misleading data signals. In the case of complex network scenarios with multiple reticulations, some of which involving transgressive events, trait estimation under tree-based models may be incorrect. On the other hand, if hybridization events are few, trait evolution is slow, and if there are no transgressive events, trait estimation might be quite close to the true value, although models may be more likely to prefer a selection-driven model like OU. In this way, it is important to explore the evolutionary conditions under which inferences are misled using tree-based models when the true evolutionary history involves reticulation. In other words, when does the phylogenetic model matter? 
Despite decades of research and methodological advancement, the specter of phylogenetic inertia in \cite{Felsenstein1985} still looms: How can we be sure that the significant results generated by our myriad methods are pointing to something meaningful rather than just a coincidence of history, or worse yet, simply an intrinsic deficiency in our method \citep{Revell2025}? 
These questions are not new and have been previously explored  in \cite{Rabosky2015}, \cite{Hahn2016}, or \cite{Matzke2022}, for instance. But comprehensive evaluation of how PCMs behave under reticulate evolutionary scenarios is not well understood. 

Here, we explore how traits generated under a reticulate evolutionary history influence the estimates of ancestral traits, evolutionary rate, and model selection using conventional tree-based PCMs.
We first comparatively evaluate these inferences when traits are generated under networks versus the major tree underlying the same network (a major tree is obtained from a network by removing each reticulation by keeping one incoming edge with the highest inheritance probability ($\gamma$; \cite{meng2009detecting}) and removing the other incoming edge followed by suppression of degree-2 nodes (i.e., nodes with one incoming and one outgoing edge)). We also explore factors that contribute the most to the error using a combination of machine learning techniques (i.e., random forest), Generalized Linear Models (GLMs), and model-based clustering. Finally, we recommend best practices and potential pitfalls of using tree-based PCMs when evidence supports potential reticulate evolutionary histories. 

\bigskip
\section*{Materials and Methods}
\begin{figure}[h!]
    \centering
    \includegraphics[width=0.85\linewidth]{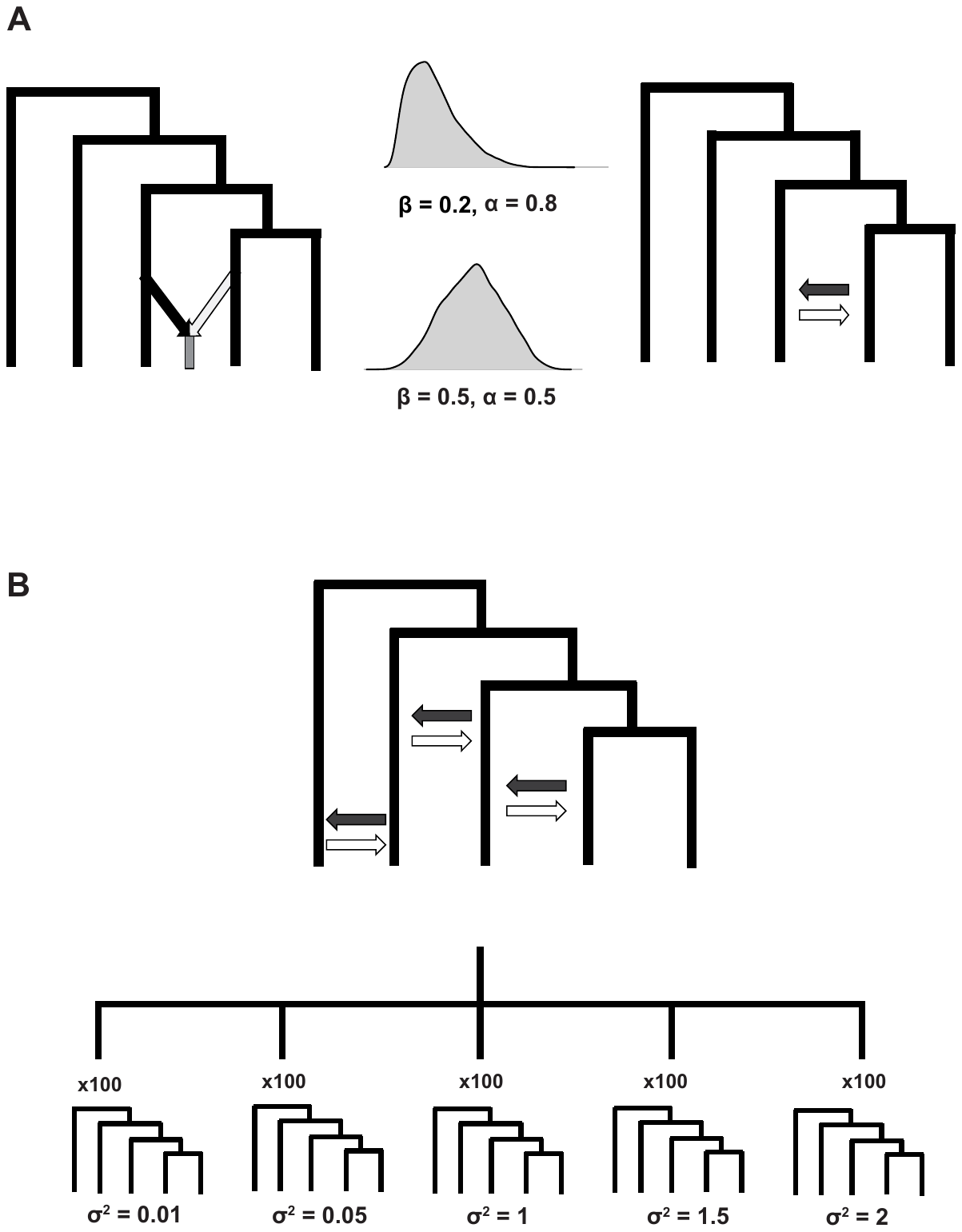}
    \caption{Visual workflow of our methodology. We began by (A) simulating species networks with various hybridization type (topologies on the left and right represent lineage-generative and lineage-neutral scenarios, respectively), and parental inheritance (drawn from one of two beta distributions in the center with $\alpha$ and $\beta$ values shown). Then, for each generated species network, (B) we simulated continuously evolving traits with 5 discrete rate parameters ($\sigma^2$), with 100 unique traits for each rate. Tip states from each character simulation were used as input for our analyses. }
    \label{fig:methods}
\end{figure}

A graphical overview of our methodology is presented in Figure \ref{fig:methods}. The aim of this study is to understand how and why tree-based PCMs may be misled under reticulate scenarios, specifically due to trait history and structural characteristics in underlying network topologies. Note all simulations and comparative modeling were performed using the true major tree of the true network, with accurate edge lengths and branching patterns, to mitigate the impacts of non-tested variables. This represents a \emph{best-case scenario}, in which researchers have access to a major tree topology that is fully accurate. We emphasize that most real-world cases will also be impacted by confounding factors, such as computational efficiency, genomic data availability, incomplete lineage sorting, and reticulation, which impact researchers' ability to accurately estimate major tree(s). 

\subsection*{Network Simulation }
We simulated a set of rooted species networks (see \cite{kong2024phynest} for a formal definition of species networks) under a birth-death-hybridization process using the \textsc{R} package \textsc{SiPhyNetworks} \citep{Justison2023SiPhyNetwork}. 
Throughout all simulations, we fixed the rates of hybridization, extinction, and speciation to 0.25, 0.2, and 1, respectively. In addition, we allowed $\gamma$ assigned to one of the edges in each reticulation to be drawn from one of two beta distribution with a mean of either $0.5$ (drawn from $\gamma\sim Beta(10,10)$ as \emph{symmetric} inheritance) or $\approx0.25$ ($\gamma\sim Beta(3,8)$ as \emph{asymmetric} inheritance).
Biologically, $\gamma$ assigned to the reticulation edge represents the proportion of genomic contribution from one parent to the hybrid descendant, which in turn, make the contribution from the other parent as $1-\gamma$. 
In this study, we make a simple yet strong assumption that $\gamma$ directly translates to the contribution of the trait from the parent, unless transgressive evolution plays a role (see subsection Trait Simulation below for detail). 
Stable inheritance represents the case where two parental species contribute roughly equally to the genomic content of the hybrid species ($\gamma\approx0.5$), whereas variable inheritance refers to the case where the contributions from the two parental species are disproportionate ($\gamma$ gets closer to zero or one).

\textsc{SiPhyNetworks} allows for three patterns of hybridization, namely lineage-generative (LG), lineage-neutral (LN), and lineage-degenerative (LD). These patterns are distinguished by the resultant change in the number of lineages immediately following a hybridization event. For example, LG hybridization results in the gain of a lineage, as in a hybrid speciation scenario; LN hybridization results in a net-zero change, as in an introgressive hybridization scenario (i.e., hybridization followed by repeated backcrossing with one of the parents, instead of hybrid lineage developing into a distinct lineage); and LD hybridization leads to the loss of a lineage following hybridization, a pattern expected, for instance, when both parental species go extinct coincident with hybridization. 

We  maximized the variability of topological patterns in the simulated species networks by using five different proportions of LG (we call this proportion $\zeta$) and LN hybridization ($\eta$) patterns during simulation using $\zeta, \eta \in \{0, 0.25, 0.5, 0.75, 1\}$, resulting in 10 combinations (Table~\ref{summary-of-sim-nets}). Note $\zeta$ and $\eta$ should sum up to one. The proportion of LD hybridization was always set to zero as the simulations resulted in too many networks that did not meet our culling threshold otherwise. For example, allowing LD hybridization often produced failed network simulations (i.e., no topology was generated) or topologies that were either trees or non-tree-based networks (we explain what we mean by `tree-based networks' below). Nevertheless, assuming the absence of LD hybridization is unlikely to pose substantial issue for the purpose our study because LN and LG hybridization are much more commonly observed pattern in current studies \citep[e.g., ][]{PardoDiaz2012,Wu2018,Banerjee2023,Owens2023}, therefore neglecting their presence will still be informative to the empiricists and our simulation indirectly mimics the LD hybridization scenarios by allowing lineages to go extinct. 

\begin{table}[]
\caption{Summary of simulated network categories. The diversity of topological patterns in the simulation networks is maximized by using a various proportions of lineage-generative ($\zeta$) and lineage-neural hybridization ($\eta$) combinations. Symmetric and asymmetric trait inheritance probabilities ($\gamma$s) represents the cases where the $\gamma$ for the reticulations is close to 0.5 and zero (or one), respectively.}
\center\begin{tabular}{l l} \hline 
Proportion of Patterns of Hybridization $(\zeta,\eta)$& Trait Inheritance Probabilities  \\ \hline 
$(1,0)$       & Symmetric\\ 
$(1,0)$       & Asymmetric\\
$(0.75,0.25)$ & Symmetric\\ 
$(0.75,0.25)$ & Asymmetric\\
$(0.5,0.5)$   & Symmetric\\ 
$(0.5,0.5)$   & Asymmetric\\
$(0.25,0.75)$ & Symmetric\\ 
$(0.25,0.75)$ & Asymmetric\\
$(0,1)$       & Symmetric\\ 
$(0,1)$       & Asymmetric\\ \hline
\end{tabular}
\label{summary-of-sim-nets}
\end{table}

Because network simulation involves stochastic processes, we simulated $>5,000$ networks, and retained a random subset of 1,000 (100 per each category in Table~\ref{summary-of-sim-nets}) that met our post-hoc criteria.
Our post-hoc criteria were designed to account for both uniformity among networks and computational efficiency. More specifically, we selected networks that had 10 leaves, contained 1--3 reticulations, and $\ge90\%$ survival rates for taxa involved in reticulate events. 
Because phylogenetic network inference remains computationally challenging, our simulations focus on small to moderately sized networks ($\approx 10$ tips), which reflect the scale of networks typically analyzed in empirical studies.
To avoid complications when using tree-based methods, we retained only simulated networks that were tree-based ({i.e.}, a phylogenetic tree with extra reticulation edge(s)). Formally, a phylogenetic network $\mathcal{N}$ on a set of taxa $X$ is tree-based if (and only if) there exists a rooted tree (i.e., support tree or also called displayed tree) that contains all vertices and a subset of the edges of $\mathcal{N}$ whose leaf set is $X$ \citep{kong2022classes}. A tree-based network can contain multiple displayed trees depending on which reticulation edge is removed to obtain the tree. 
Among these, we selected the \emph{major tree}, which is the one with the highest weight, for use in our downstream analyses. This weight is computed by multiplying all of the $\gamma$s assigned to the edges that are not removed in the production of the major tree. 
Note all tree edges have $\gamma=1$ whereas all reticulate edges have $\gamma=(0,1)$.
For each simulated network, we recorded details of relevant network structural characteristics.
See Tables \ref{vars_Description} and \ref{Vars_Description_NS} for all collected variables.

\subsection*{Trait Simulation}

For each simulated rooted species network with all topological parameters ({i.e.}, edge lengths and $\gamma$s), we used the \textsc{Julia} package \textsc{PhyloNetworks} version 0.16.4 \citep{SolisLemus2017} to simulate a continuous trait at every vertex, specifying that traits evolve under BM over time. 
We set the the starting value of the trait (ancestral mean) to zero across all simulations and used variance (the evolutionary rate parameter) values of $\sigma^2\in\{0.1, 0.5, 1.0, 1.5, 2.0\}$. We will refer to these as \emph{network-based simulations} hereafter. 
Traits generated with small variance values are biologically interpreted as slow-evolving, while those generated with larger variance values are expected to exhibit rapid evolution. More precisely, let us call the trait value of a tree node $v$ as $Y_v$, whose parent is node $u$ that has the trait value of $Y_u$. The trait value $Y_v$ is drawn from the normal distribution with mean of $Y_u$ and with variance $\sigma^2\ell_e$ where $\ell_e$ is the edge length of $e$ from $u$ to $v$. In contrast, for a hybrid node $h$ with the parent nodes $a$ and $b$ (with the trait value of $Y_a$ and $Y_b$, respectively), normal distribution is assumed for the trait value of $h$ ($Y_h$) where the mean is ${\gamma_e}_aY_a+{\gamma_e}_bY_b$ where $\gamma$ is the inheritance probability and is assigned to $e_a$ and $e_b$, the two reticulation edges that start from $a$ and $b$, respectively, toward $h$. Note ${\gamma_e}_b$ is $1-{\gamma_e}_a$ if the network is assumed to be binary and every hybrid node has indegree two (i.e., two parental species to a hybrid). In case of time-consistent reticulation (i.e., both $e_a$ and $e_b$ have length zero), $Y_h$ has variance of zero. Otherwise, the conditional variance of $Y_h$ is ${\gamma_e}_a^2\sigma^2\ell_{e_a}+{\gamma_e}_b^2\sigma^2\ell_{e_b}$ \citep{Bastide2018}. In other words, the concept of additivity, where the phenotypic characteristics of the hybrid offspring are a simple average of the parental traits weighted by $\gamma$, is generally assumed in our simulation.

To improve the realism of the trait simulation, we additionally accommodated transgressive evolutionary scenarios. Transgressive evolution describes the process in which hybrid offspring exhibit phenotypic traits that exceed the limits of their parental traits, contradicting the concept of additivity. In terms of trait simulation, accommodating transgressive evolution can be seen as a similar concept to the shifted BM \citep{bastide2018inference}, where the entire path is shifted by a constant value while still maintaining the same random fluctuations. In our simulation, such a shift is applied to the starting point of the edge directly following the hybridization event (i.e., the edge that has the reticulation vertex as the parent). We set the probability of hybridization-driven transgressive evolution at a reticulation node to $0.1$. While the frequency of transgressive phenotypes in hybrid populations in reality may be higher, particularly in the cases of domesticated, intraspecific crosses \citep{Rieseberg1996}, we believe our setting is appropriate for the purpose of our study. For the trait at the reticulation vertex selected to represent transgressive evolution, the trait value $x$ is exaggerated by adding a value $\delta$  drawn from a randomly selected one of the following two uniform distributions: $\delta \sim \mathcal{U}(-20\sigma^2,~x-\sigma^2)$ and $\delta \sim \mathcal{U}(x+\sigma^2,~20\sigma^2)$. 
In other words, we allow a transgressive event to result in either a decrease or an increase in the trait value at random, with the magnitude of the change ranging between the expected value under no transgressive effect (i.e., $\sigma^2$) and 20 times that value.
Otherwise, the trait values of the descendants of hybridization follow the additivity rule at probability of 0.9. One hundred replicates were made for each $\sigma^2$ value.

\subsection*{Trait and Parameter Estimation}

We used the \textsc{R} packages \textsc{phytools} \citep{Revell2012} and \textsc{geiger} \citep{Harmon2008} to conduct ACE, estimate trait evolutionary rates, and obtain the best-fit models of trait evolution (between BM and OU) using AIC. 
We restrict ACE to tree vertices in the major tree of the network of interest because these methods only assume trees as the underlying model. Using the major trees and tip states, we first select either BM or OU by evaluating model fit using AIC. Second, using the model selected (as well as the major trees and tip states), we perform ACE to obtain both an estimated rate parameter and estimated trait values at all internal tree nodes. 
Here, we know the true model under which traits evolved (BM), the true rate at which they evolved ($\sigma^2$), and the true trait states at each ancestral node. Thus, we are able to assess error in each of these three essential ACE processes.
We measured trait estimation error for each node, then computed average tree-wide trait estimation error for each network- and tree-based simulations (see next section for a definition). We also measured evolutionary rate estimation error by comparing the estimated value with the true value used for simulation. Hereafter, we will use ``mean tree-wide trait estimation error'' to refer to standardized mean error in ancestral trait estimations across a tree, ``node-specific trait estimation error'' to refer to standardized error in ancestral trait estimations at specific tree nodes,  and ``rate estimation error'' to refer to error in evolutionary rate estimations. 

\subsection*{Comparing Estimates with Trees}

It is important to note that even when we observe substantial errors in ACE and rate estimation, such deviations may not be due to model violations, but rather to limitations of the methods implemented in \textsc{phytools} and \textsc{geiger}. To clearly demonstrate that the observed error fell outside the expected error range when classic bifurcating phylogenies describe trait evolution, we additionally simulated continuously evolving traits under classic tree-based models. This procedure can be considered analogous to a negative control. 
We randomly selected with replacement 2,000 major trees from our set of generated networks, then we used \textsc{PhyloNetworks} to simulate characters evolving under BM with identical rate values used in our network-based trait simulations.
This procedure resulted in 100,000 simulated continuous traits, which we will refer to as \textit{tree-based simulations} hereafter.
For each dataset generated in the tree-based simulation, we followed the same procedure as in the network-based simulation described above.
To compare tree-based results to network-based results, we used two-tailed t-tests to test for significant differences among mean rate and trait estimation error distributions. We used Levene's test statistic to asses homoscedasticity (i.e., equal distribution of variance between two groups) among tree-based and network-based error distributions. To test for significant differences among model choice, we employed one- and two-proportion Z-tests to assess the null hypotheses that the proportion of incorrectly chosen models was equal between tree- and network-based simulations, and that the proportion of incorrectly chosen models were not higher in network-based simulations, respectively. 

\subsection*{Factors Contributing to Error}

Structural characteristics of network topology and trait history therein collected during simulations were used to determine which factors contribute most strongly to observed error, and which combinations of trait evolution and network structure lead to the largest error. 
Tables \ref{vars_Description} and \ref{Vars_Description_NS} describe all collected variables for topology-wide and node-specific downstream analyses, respectively.

\begin{table}
\caption{Description of all variables related to network topology and trait history collected in this study, and their identifiers used to analyze factors contributing to estimation error.}
\label{vars_Description}
\begin{threeparttable}
\begin{tabular}{l p{0.75\linewidth}}
\hline
Variable name & Description \\ \hline
\texttt{sig2\_True} & True evolutionary rate of the trait \\
\texttt{ModelUsed} & Best-fit model of trait evolution selected by AIC \\
\texttt{numTrans} & Number of transgressive events \\
\texttt{termMax} & Maximum terminal branch length \\
\texttt{termMin} & Minimum terminal branch length \\
\texttt{termMean} & Mean terminal branch length \\
\texttt{termVar} & Variance in terminal branch lengths \\
\texttt{intMax} & Maximum internal branch length \\
\texttt{intMin} & Minimum internal branch length \\
\texttt{intMean} & Mean internal branch length \\
\texttt{intVar} & Variance in internal branch lengths \\
\texttt{termRat} & Ratio of mean internal to mean terminal branch length \\
\texttt{nHyb} & Number of hybridization events (1, 2, or 3) \\
\texttt{PropTermH} & Proportion of reticulation nodes occurring on terminal branches \\
\texttt{PropExtH} & Proportion of reticulation nodes occurring on extinct branches \\
\texttt{HybType} & Probability of a reticulation following lineage generative (LG) or lineage neutral (LN) hybridization pattern\tnote{1} \\
\texttt{inherMean} & Mean inheritance values \\
\texttt{inherMax} & Maximum inheritance probabilities \\
\texttt{inherVar} & Variance in inheritance probabilities \\ \hline
\end{tabular}
\begin{tablenotes}[flushleft]
\footnotesize
\item[1] LG = 100\% lineage generative; LN = 100\% lineage neutral; LG75 = 75\% lineage generative and 25\% lineage neutral; LN75 = 75\% lineage neutral and 25\% lineage generative; LG50 = 50\% lineage generative and 50\% lineage neutral.
\end{tablenotes}
\end{threeparttable}
\end{table}

\begin{table}
\caption{Description of all variables collected in this study and their identifiers used to analyze factors contributing to node-specific trait estimation error.}
\label{Vars_Description_NS}
\begin{threeparttable}
\begin{tabular}{l p{0.75\linewidth}}
\hline Variable Name & Description \\ \hline
\texttt{NodeDepth} & Branch-length adjusted node depth by setting root=0 \\
\texttt{MinDistH} & Minimum absolute distance of node to a hybridization event, measured in branch length units\\
\texttt{MinPosDistH} & Minimum positive distance of a node to a hybridization event\tnote{1}\\
\texttt{MinNegDistH} & Minimum negative distance of a node to a hybridization event\tnote{2} \\
\texttt{TransMagnitude} & Magnitude of transgressive evolution at closest reticulation to a node \\
\texttt{brLenFrom} & Mean branch length immediately following a focal tree node\\
\texttt{brLenTo} & Mean branch length immediately leading to a focal tree node\\ 
\hline
\end{tabular}
\begin{tablenotes}[flushleft]
\footnotesize
    \item[1] Node occurs following a hybridization event; \item[2] Node occurs in prior to a hybridization event
\end{tablenotes}
\end{threeparttable}
\end{table}

We used a combination of machine learning techniques (i.e., random forest), Generalized Linear Models (GLMs), and model-based clustering to understand which factors were the best predictors of observed error in trait estimation, rate estimation, and model choice.
Random forest is a non-parametric machine-learning technique capable of modeling complex interactions and providing robust measures of variable importance across a wide range of predictor combinations.
GLMs offers a more interpretable and statistically grounded framework for testing specific hypotheses about how individual predictors contribute to estimation error. 
Finally, Model-based clustering allowed us to identify latent structure in the data without specifying groupings in prior, which helps to reveal natural clusters of conditions associated with high or low error. To determine what factors contribute to model choice, a model-based clustering method implemented in the \textsc{R} package \textsc{mlclust} \citep{Scrucca2016mclust} was used to classify BM versus OU model choice situations according to explanatory variables. 

\bigskip
\section*{Results}\label{sec3}
\subsection*{Trait Estimation Error}
\subsubsection*{Tree- versus network-based evolution}
\begin{figure}[h!]
    \centering
    \includegraphics[width=1\linewidth]{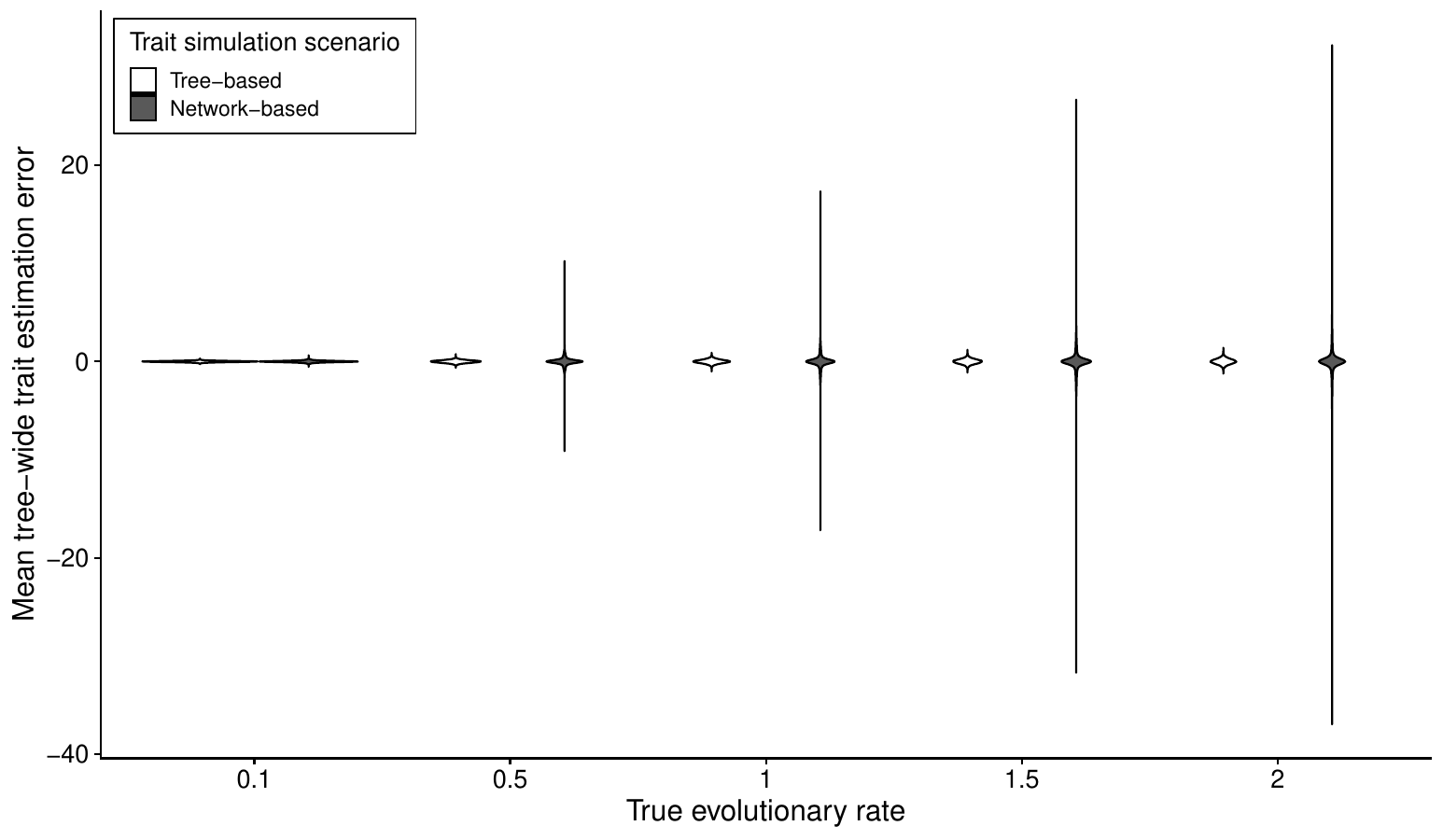}
    \caption{Comparison of mean tree-wide trait estimation error (y-axis) between network-based (dark gray; right violin for each true evolutionary rate) and tree-based (white; left violin for each true evolutionary rate) simulations across true evolutionary rates (x-axis), when a tree-based phylogenetic comparative method was used for estimation. Values are shown on the original scale rather than a log-transformed scale to preserve tail behavior, highlighting the increasing variance in estimation error for network-based simulations as the true evolutionary rate increases.}
    \label{fig:Tree-vs-Network-Est}
\end{figure}

Figure~\ref{fig:Tree-vs-Network-Est} presents that variance in mean tree-wide trait estimation error increases with a trait's true evolutionary rate for both tree- and network-based simulations. However, this increase in variance is substantially more pronounced for network-based simulations. On average, tree-based PCMs led to a significant increase (two-tailed t-test, $p\approx0.0005$) in overall error for network-based simulations compared to tree-based simulations. More specifically, mean overall trait estimation error for network-based simulations was $4.1\times10^{-3}$, while it was $-2.0\times10^{-4}$ for tree-based simulations. The two datasets exhibited significant heteroscedasticity (Levene's test, $p<2.2\times10^{-16}$), indicating that while network-based scenarios may not severely hamper accuracy, they do significantly impact precision. On the other hand, Kruskal–Wallis tests did not detect significant differences in central tendency between the two groups.

\subsubsection*{Factors contributing to mean tree-wide trait estimation error}
\begin{figure}[h!]
    \centering
    \includegraphics[width=1\linewidth]{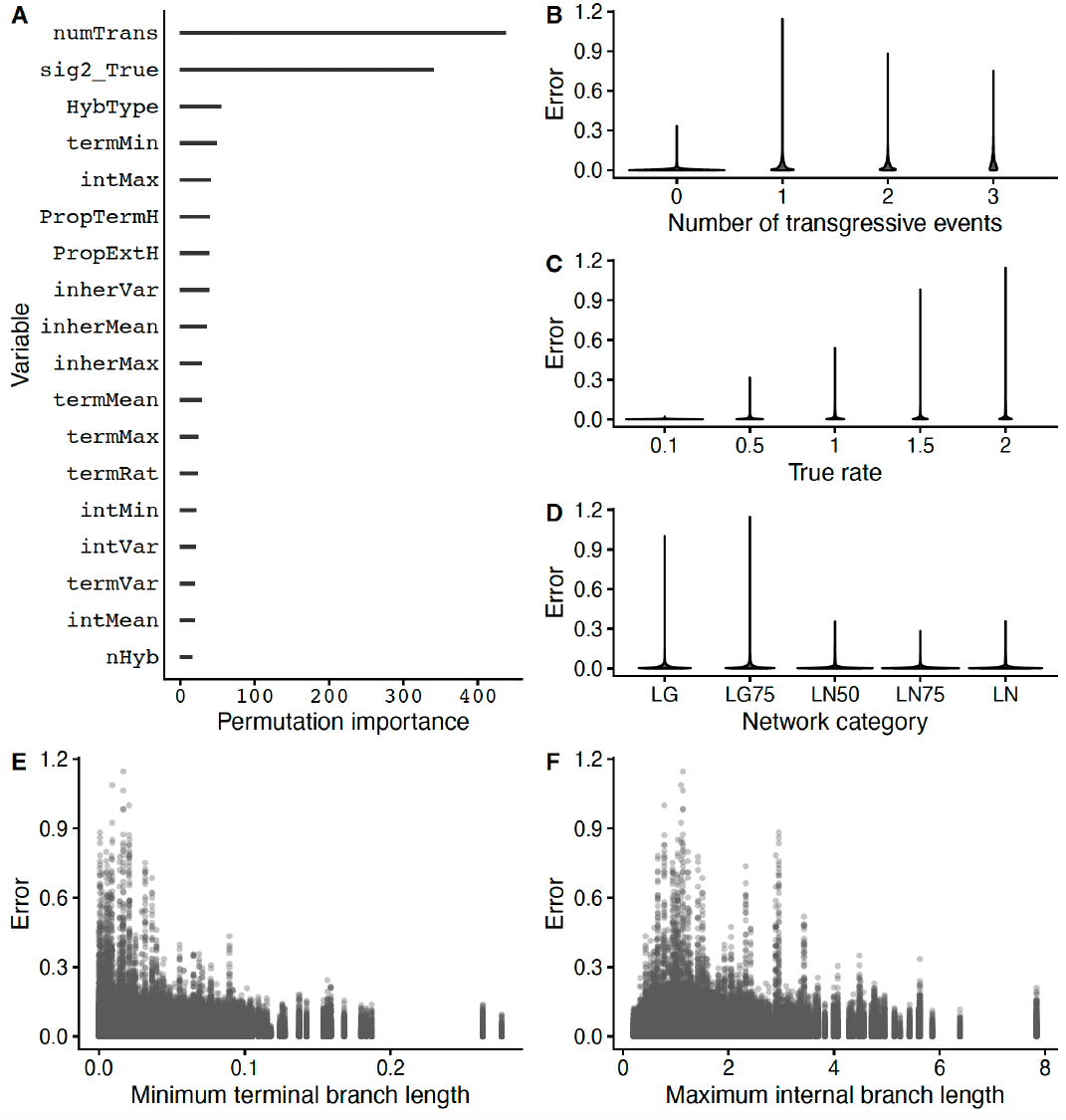}
    \caption{(A) Permutation importance of factors contributing to variation in tree-wide trait estimation error under network-based simulations from random forest analyses. In (B)–(F), error (y-axis) refers to mean tree-wide trait estimation error (log-transformed). In (B)–(D), we show estimation error for the three variables identified as most important in (A): (B) number of transgressive events, (C) values of the true evolutionary rate, and (D) network categories (see Table~\ref{vars_Description} for acronyms). In (E) and (F), we show that shorter branch lengths contributed to increased tree-wide trait estimation error.}
    \label{fig:factors-contributing-to-tree-Est}
\end{figure}

Figure~\ref{fig:factors-contributing-to-tree-Est} summarizes our findings regarding factors associated with mean tree-wide trait estimation error.
Random forest analyses using all predictor variables (as listed in Table~\ref{vars_Description}) explained 50\% of observed variance (Fig.~\ref{fig:factors-contributing-to-tree-Est}A). 
The most important variables were the number of transgressive events (\texttt{numTrans}) and the true evolutionary rate of the trait (\texttt{sig2\_True}), followed by the probability of a reticulation following the LG or LN pattern (\texttt{HybType}).
Models run with the first two variables alone explained 34\% of observed variance. 
Figures~\ref{fig:factors-contributing-to-tree-Est}B--\ref{fig:factors-contributing-to-tree-Est}D show log-transformed mean tree-wide trait estimation error for different values of these three variables.
Interestingly, mean error for the networks with LN pattern was smaller relative to LG pattern (Fig.~\ref{fig:factors-contributing-to-tree-Est}D). 
Branch lengths also influenced mean tree-wide trait estimation accuracy. Specifically, overall error decreased as maximum internal branch lengths and minimum terminal branch lengths increased (Figs.~\ref{fig:factors-contributing-to-tree-Est}E and~\ref{fig:factors-contributing-to-tree-Est}F).
This can be interpreted to mean that the presence of short terminal branches (represented by minimum branch lengths) increases error, while the presence of longer internal branches (represented by maximum branch lengths) decreases error.  

\begin{figure}[h!]
    \centering
    \includegraphics[width=1\linewidth]{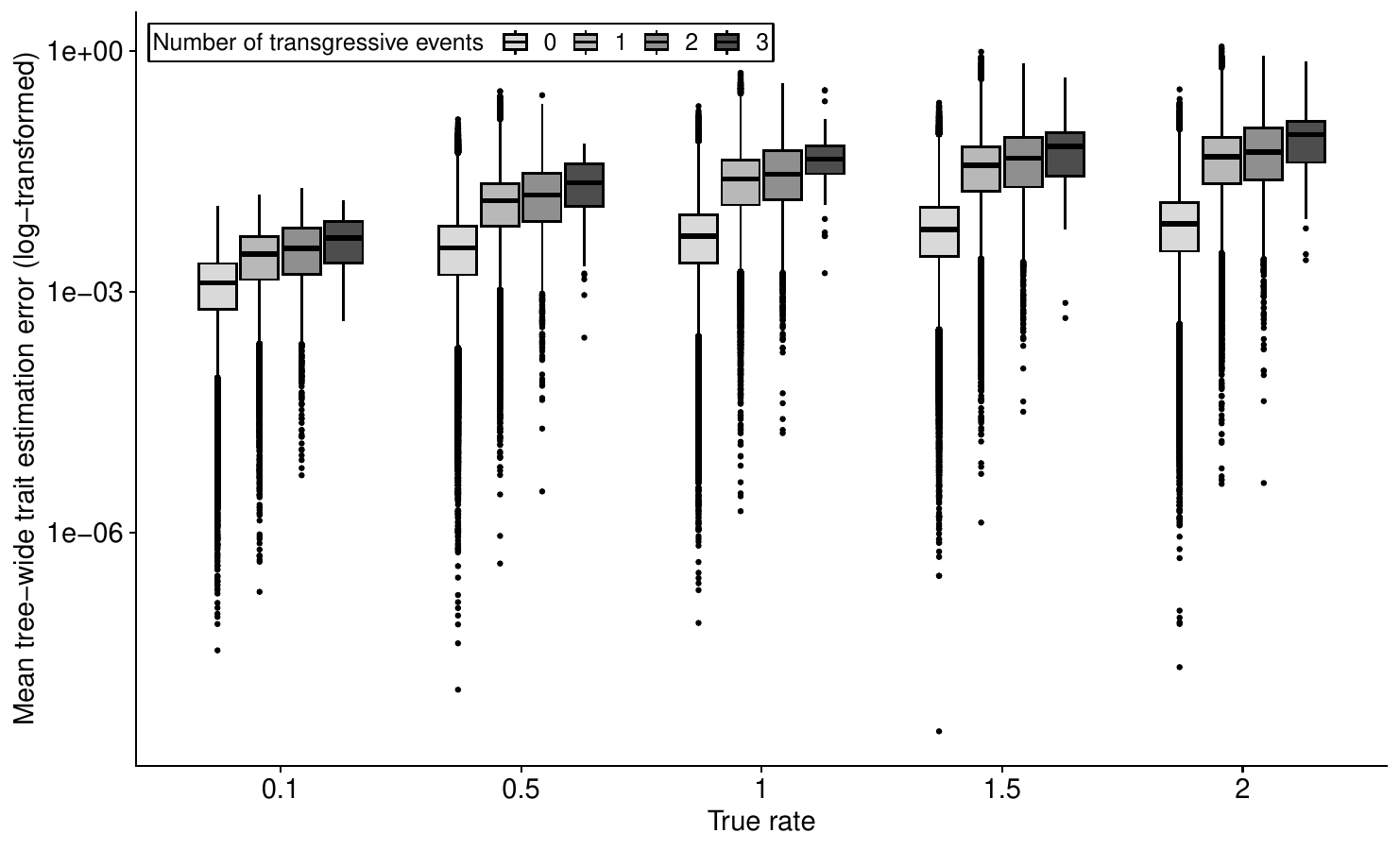}
    \caption{Combined effects of evolutionary rate and transgressive evolution on mean tree-wide estimation error (y-axis; log-transformed). As trait evolutionary rates increase, the effects of transgressive evolution become more pronounced.}
    \label{tree-wide-transRate-int}
\end{figure}

Generalized linear models revealed that a combination of trait-history and network structure features contribute to observed increases in trait estimation error for the reticulate scenarios. 
The best-fit model (i.e., BM or OU), which predicted absolute estimation error, had a small but significant pseudo-$R^2$ of $0.284$. 
Also, GLMs and 3-way ANOVA revealed significant positive interaction effects among variables identified as important in random forest analysis, indicating that true evolutionary rate interacts agonistically with transgressive evolution.
In other words, the effects of transgressive evolution on character displacement become more pronounced as traits evolve more rapidly (Fig.~\ref{tree-wide-transRate-int}), leading to increased mean tree-wide trait estimation error.
This is because character displacement is more likely to result in larger shifts when the true rate of trait evolution is higher.

\subsubsection*{Factors contributing to node-specific trait estimation error}
\begin{figure}[h!]
    \centering
 \includegraphics[width=1\linewidth]{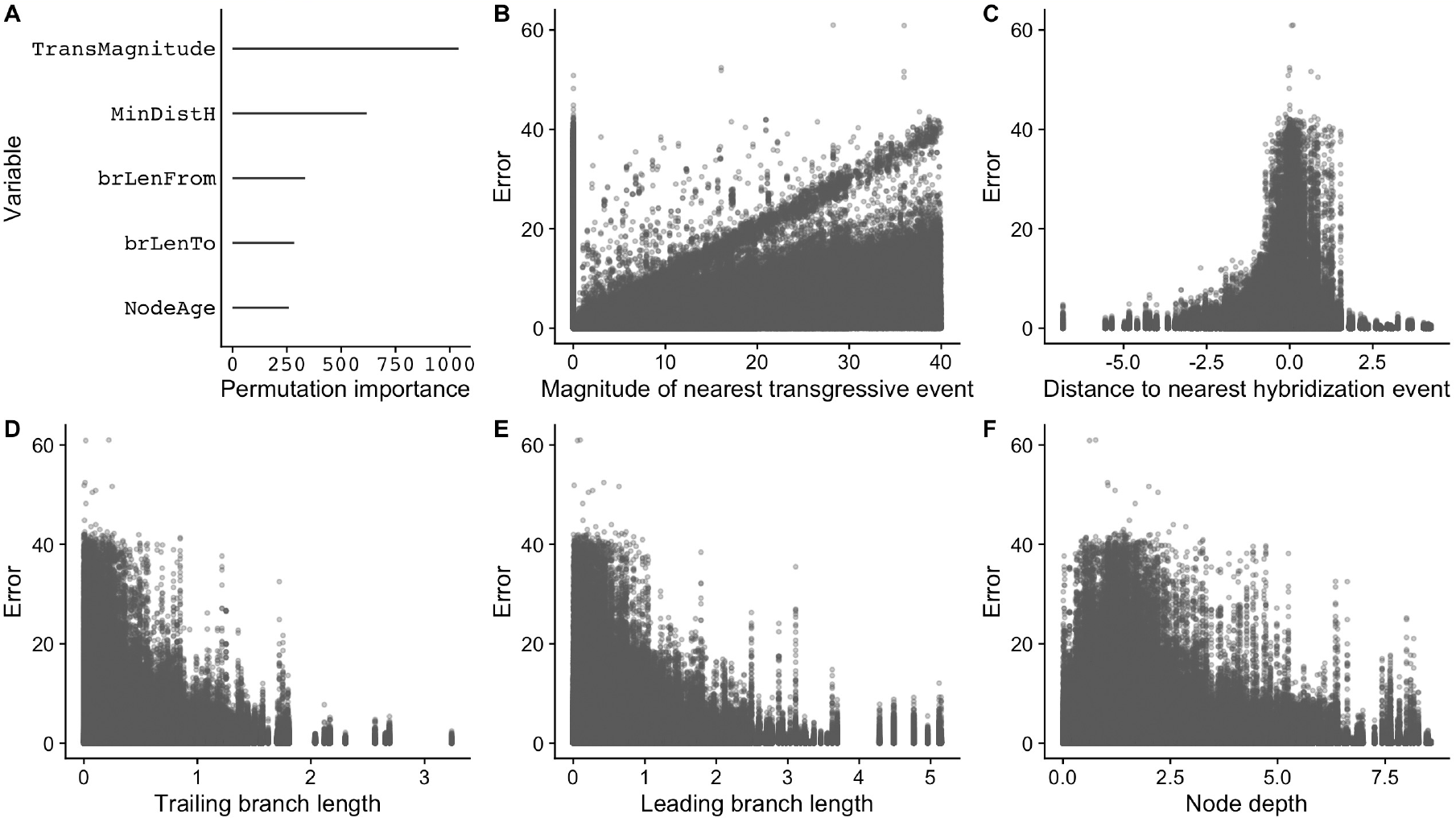}
    \caption{(A) Permutation importance from random forest analyses of factors influencing node-specific trait estimation error under network-based simulations. The most important variables were (B) magnitude of nearest transgressive event and (C) distance to the nearest hybridization event. (D--E) Nodes on shorter branches experienced greater error. Trailing and leading branch lengths refer to the mean branch lengths immediately before and after a focal tree node, respectively. (F) Older nodes showed higher error. In (B)--(F), error (y-axis) refers to mean node-specific trait estimation error.}
    \label{fig:node-specific}
\end{figure}

Figure \ref{fig:node-specific} summarizes our results regarding node-specific trait estimation errors for network-based simulations that used tree-based PCMs.
Random forest analyses using the variables listed in Table \ref{Vars_Description_NS} explained 42\% of observed variance. 
The two most important variables were the magnitude of transgressive evolution (\texttt{TransMagnitude}) and the distance to the nearest hybridization event (\texttt{MinDistH}), followed by branch lengths from (\texttt{brLenFrom}) and to (\texttt{brLenTo}) reticulation nodes, and node age (\texttt{NodeAge}) (Fig.~\ref{fig:node-specific}A). 
The magnitude of the nearest transgressive evolution event and the distance to the nearest hybridization event from the node of interest explained 25\% of the total variance in node-specific trait estimation error (Fig.~\ref{fig:node-specific}B--\ref{fig:node-specific}C). 
Interestingly, nodes closest to hybridization events tend to suffer from the highest error rates regardless of whether they preceded or followed them (Fig. \ref{fig:node-specific}C). 
Thus, increased node-specific estimation error is expected at nodes near reticulation events that resulted in transgressive evolution. Generalized linear models showed very low but significant pseudo-$R^2$ of $<0.1$, so these models were not considered. 

Nodes on short branches (whether preceding or following the focal node) show increased estimation error (Figs.~\ref{fig:node-specific}D--\ref{fig:node-specific}E). This suggests that short intervals between speciation events (e.g., adaptive radiation) reduce estimation accuracy, which is expected to be pronounced under reticulate scenarios. 
Finally, nodes closer to the root (node depth $\approx 0$) tended to have higher estimation error (Fig. \ref{fig:node-specific}F). 

\subsection*{Rate Estimation Error}
\subsubsection*{Tree- versus network-based evolution}

\begin{figure}[h!]
    \centering
    \includegraphics[width=\linewidth]{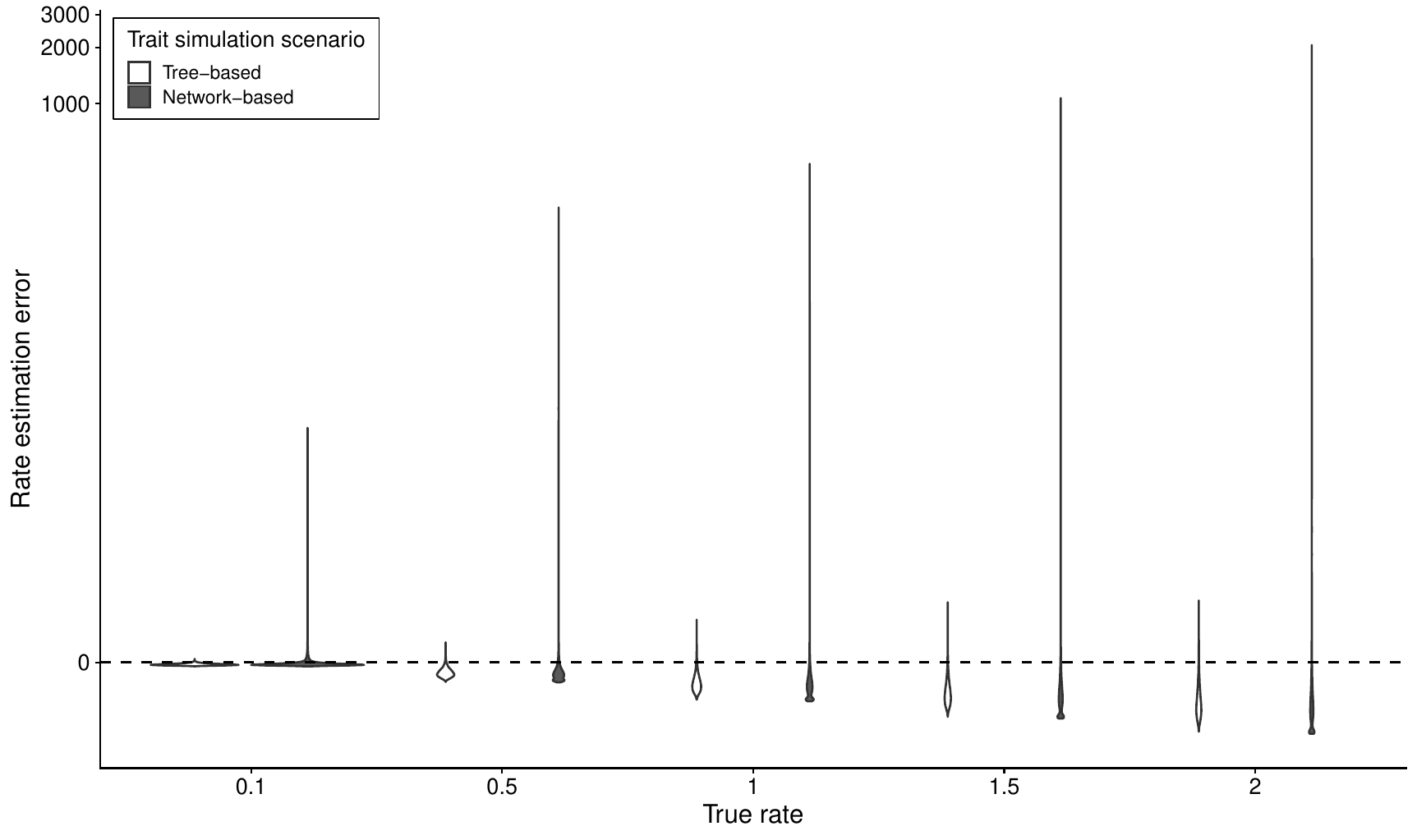}
    \caption{Rate estimation error (y-axis) for tree-based (white; left violin for each true rate) and network-based (dark gray; right violin for each true rate) simulations across true evolutionary rates. The dotted horizontal line indicates perfect estimation; values above and below the line represent over- and underestimation, respectively. Kruskal–Wallis test indicate significant differences in mean error among groups.}
    \label{Tree-vs-Network-rate-Est}
\end{figure}

Figure~\ref{Tree-vs-Network-rate-Est} presents rate estimation error for tree- and network-based simulations. Mean overall rate estimation error was significantly different (two-tailed t-test, $p<2.2\times10^{-16}$) between network- (mean = $1.8812$) and tree-based simulations (mean = $-0.5078$).  

As the true rate of evolution ($\sigma^2$) increased, differences in mean rate estimation error and estimation variance also increased, with network-based simulations experiencing increasing overestimation (mean = $5.3330$) and tree-based simulations experiencing increasing underestimation (mean = $-0.9872$). When traits evolved slowly ($\sigma^2=0.01$), however, network-based simulations saw smaller mean estimation error (mean = $-0.0206$) than tree-based simulations (mean = $-0.0507$). 
Similar to tree-wide trait estimation comparisons, distributions exhibited significant heteroscedasticity (Levene’s test, $p<2.2\times10^{-16}$), indicating that reticulate evolutionary histories impact both accuracy and precision of rate parameter estimations, with reticulate scenarios leading to significant overestimation of the rate parameter in general. 

\subsubsection*{Factors contributing to rate estimation error}
\begin{figure}[h!]
    \centering
    \includegraphics[width=1\linewidth]{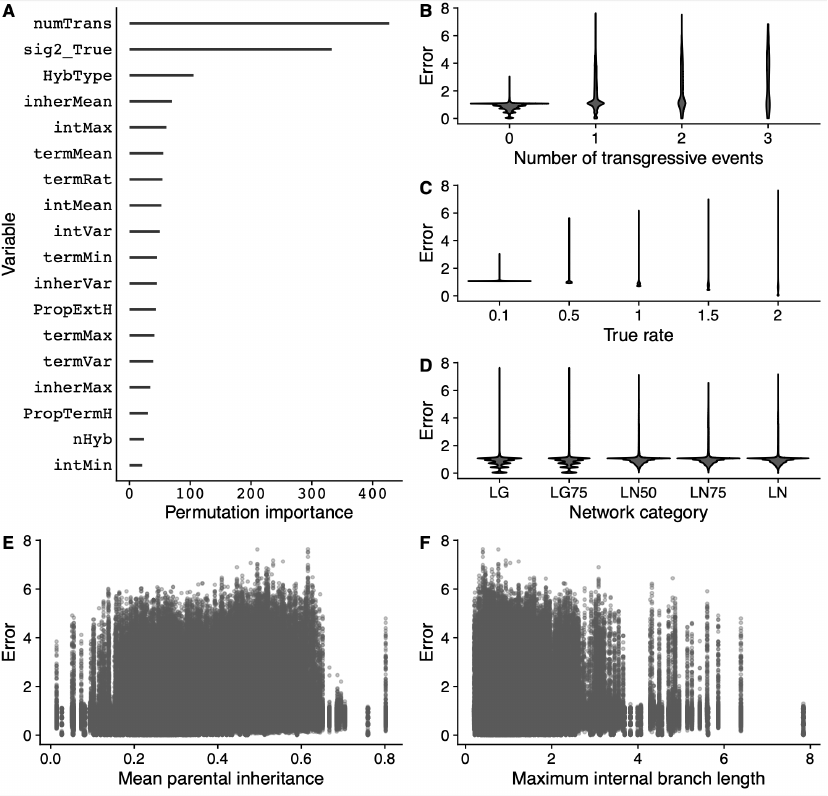}
    \caption{(A) Permutation importance of factors contributing to variation in rate estimation error under network-based simulations from random forest analyses. In (B)--(F), errors (y-axis) refers to rate estimation error. All rate estimation errors were log-transformed for plotting for clarity, where zero represents the minimum rate estimation error. In (B)--(D), we show rate estimation error for the three variables identified as most important: (B) number of transgressive events, (C) values of the true rate, and (D) network categories (see Table~\ref{vars_Description} for acronyms). (E) Rate estimates were least accurate when parental inheritance values were symmetrical, and (F) shorter internal branch lengths contributed to increased rate estimation error.  }
    \label{fig:factors-rate-est}
\end{figure}

Figure \ref{fig:factors-rate-est} summarizes our results regarding evolutionary rate estimation errors for network-based simulations that used tree-based PCMs.
Random forest analyses using the full set of predictor variables (Table~\ref{vars_Description}) explained 77\% of the observed variance.
The most important factors determining error were the number of transgressive events (\texttt{numTrans}), the true rate of trait evolution (\texttt{sig2\_True}), the proportion of hybridization scenarios (\texttt{HybType}), and mean parental inheritance (\texttt{inherMean}) (Fig.~\ref{fig:factors-rate-est}A). These four variables could explain 60\% of observed variance. Variables identified as important in GLM were in concordance with the random forest analysis.
Figures~\ref{fig:factors-rate-est}B--\ref{fig:factors-rate-est}D show rate estimation error for different values in the three important variables.
Further, symmetrical mean parental inheritance negatively influenced to the rate estimation (Fig.~\ref{fig:factors-rate-est}E) and short internal branches increase rate estimation error (Fig.~\ref{fig:factors-rate-est}F).

\begin{figure}[h!]
    \centering
    \includegraphics[width=\linewidth]{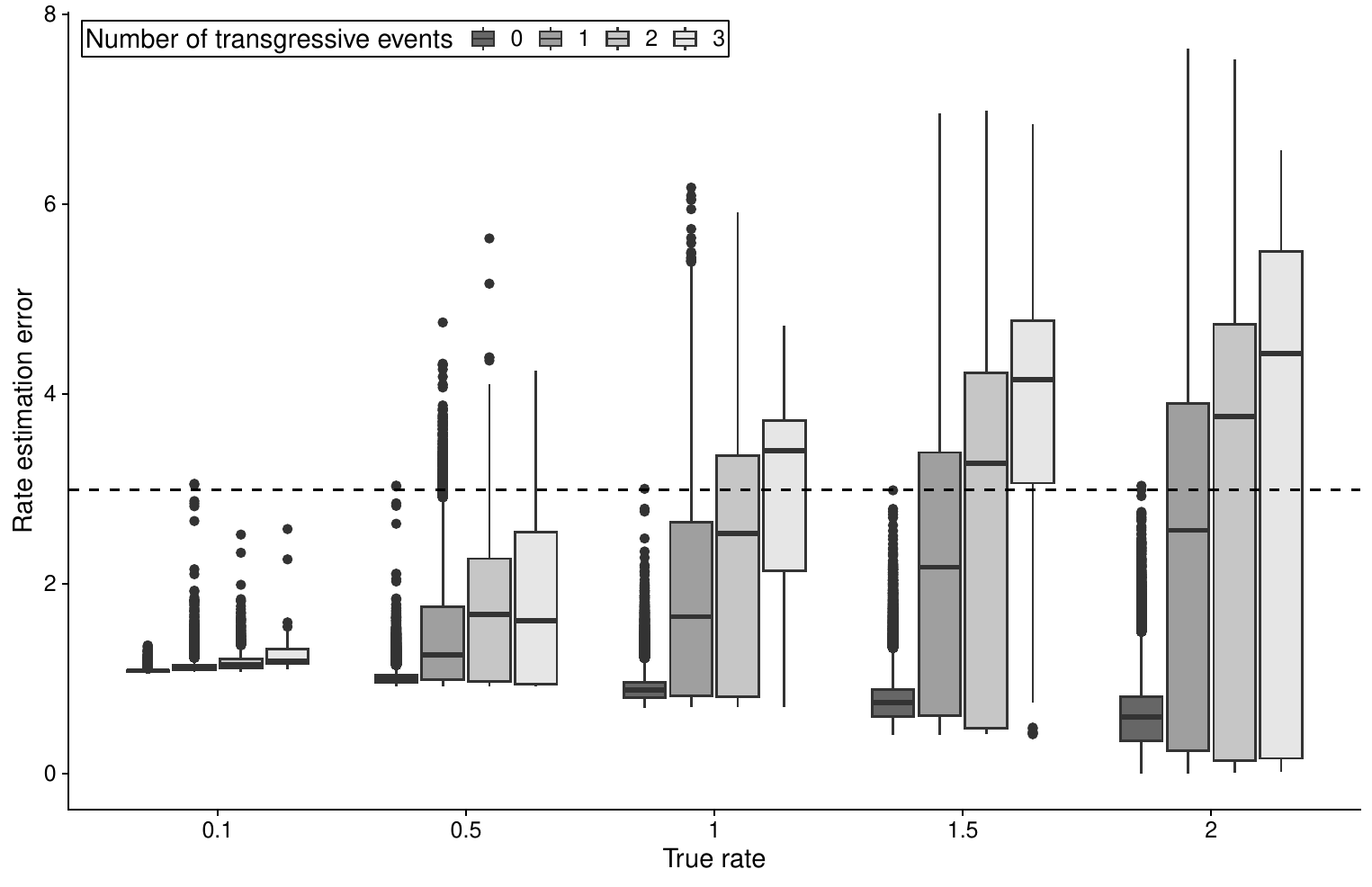}
    \caption{Combined effects of rate and transgressive evolution on rate estimation error (y-axis). X-axis represents the true evolutionary rate used for simulation. Values were standardized to enhance visualization. The dotted horizontal line at $y=3$ represents perfect estimation. Areas above and below the dotted horizontal line indicate rate overestimation and underestimation, respectively. As traits evolve more rapidly, the effects of transgressive evolution are magnified, leading to increased variance, and increased chances of rate overestimation. Without transgressive evolution, rates tend to be underestimated in reticulate scenarios.}
    \label{fig:Rate-transRate-int}
\end{figure}

Interaction effects between true rate and number of transgressive events were significant ($p<2.2\times10^{-16}$) and positive, beginning at a $\sigma^2=0.5$. 
Under GLMs, interaction effects ranged from 0.3898 ($\sigma^2 = 0.5$; single transgressive event) to 2.7596 ($\sigma^2 = 2$; three transgressive events). This effect magnitude increased with both $\sigma^2$ and the number of transgressive events.
Furthermore, as the true evolutionary rate increases, the effects of transgressive evolution become more pronounced (Fig.~\ref{fig:Rate-transRate-int}), resulting in larger magnitudes of both overestimation and underestimation.
As the evolutionary rate increase, trait displacement resulting from transgressive evolution becomes more pronounced, leading to rate overestimation. In contrast, in the absence of transgressive evolution, reticulation homogenizes variation, leading to rate underestimation that becomes more pronounced as traits evolve more rapidly. Despite this, the magnitude of rate overestimation outweighed the magnitude of rate underestimation across the board. 

\subsection*{Model Choice}
\subsubsection*{Tree- versus network-based evolution}
Table~\ref{table:bm-and-ou} shows the proportion of selected models for traits generated under a network versus a tree. Brownian Motion is generally correctly favored in both scenarios. Based on AIC, OU was incorrectly favored (since all traits were generated under BM) in $\approx21\%$ of the cases evaluated for network-based simulations. In contrast, the frequency of such false selection was substantially lower ($\approx7\%$) when the model selection was performed for tree-based simulations. 
This difference was statistically significant (one-tailed Z-test, $p<2.2\times10^{-16}$), indicating that features of reticulate evolution may make interpretations from best-fit models of evolution a dubious endeavor. 

\begin{table}[h!]
\caption{Proportion of models in which Ornstein–Uhlenbeck (OU) or Brownian Motion (BM) were favored by model comparison statistics for tree- and network-based simulations. BM is generally favored as the correct model}
\center\begin{tabular}{l l l} \hline 
Scenario & Proportion of BM selected & Proportion of OU selected  \\ \hline 
Tree-based    &   93\%    & 7\% \\ 
Network-based &   79\%& 21\%\\ \hline 
\end{tabular}
\label{table:bm-and-ou}
\end{table}

\subsubsection*{Factors contributing to differences in model choice}
\textsc{mlclust} was used to build a classification model to identify the variables (Tables~\ref{vars_Description}--\ref{Vars_Description_NS}) that favor OU or BM as the underlying model of trait evolution.
For clarity, we refer to the classification model simply as the `model', and denote ACEs in which OU or BM was selected as the best-fitting model of character evolution as the `OU group' and `BM group', respectively.
Our model had an overall classification error $\approx0.21$, with the OU group being misclassified more frequently than the BM group, which we suspect is due to differences in sample size.

\begin{figure}[h!]
    \centering
    \includegraphics[width=1\linewidth]{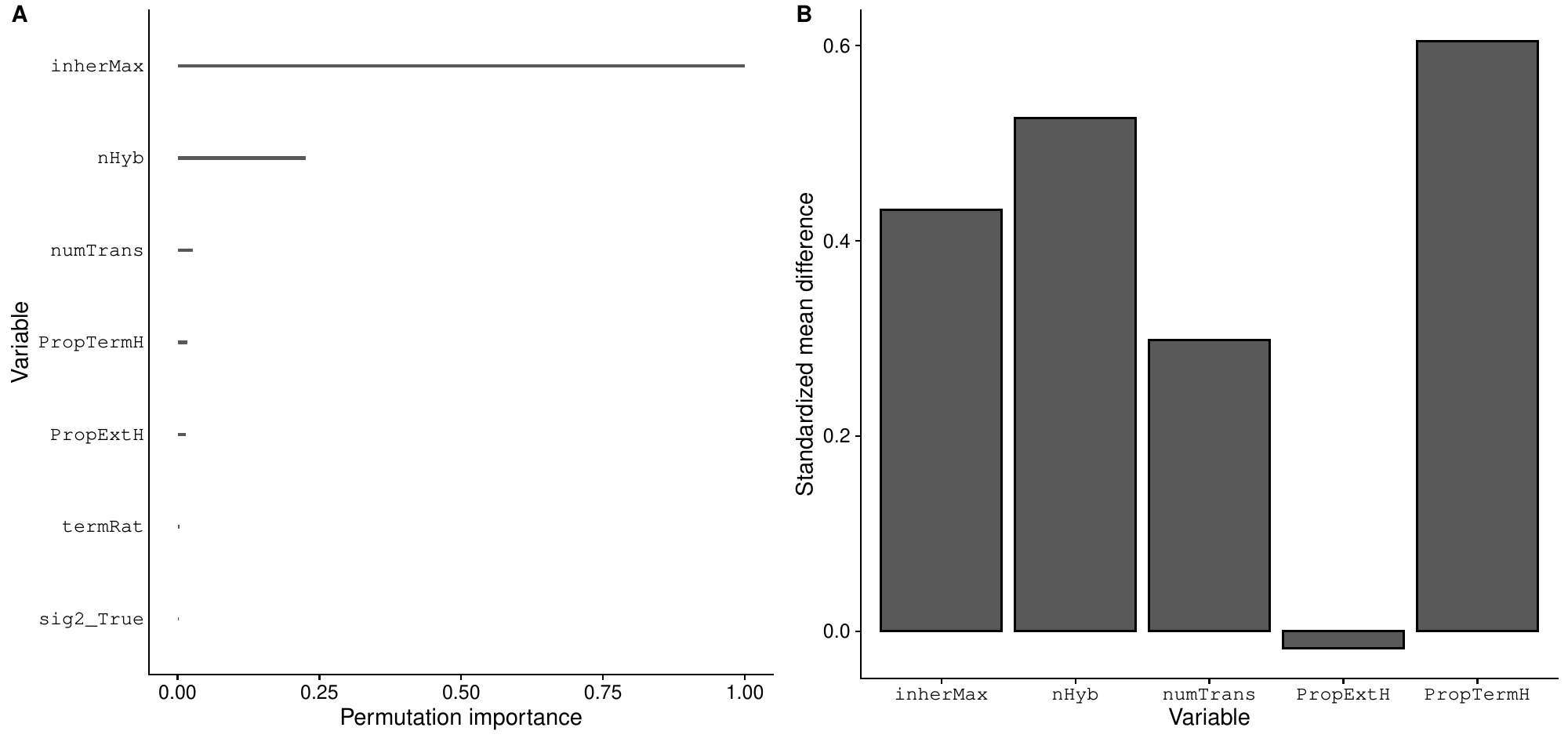}
    \caption{(A) Permutation importance of variables in \textsc{mclust} classification of Ornstein–Uhlenbeck (OU) versus Brownian Motion (BM) groups. (B) Comparison of standardized mean differences among the variables identified as important. Positive values indicate higher means in the OU group, while negative values indicate higher means in the BM group.}
    \label{fig:variables-bm-vs-out}
\end{figure}

\begin{figure}[h!]
    \centering    
    \includegraphics[width=1\linewidth]{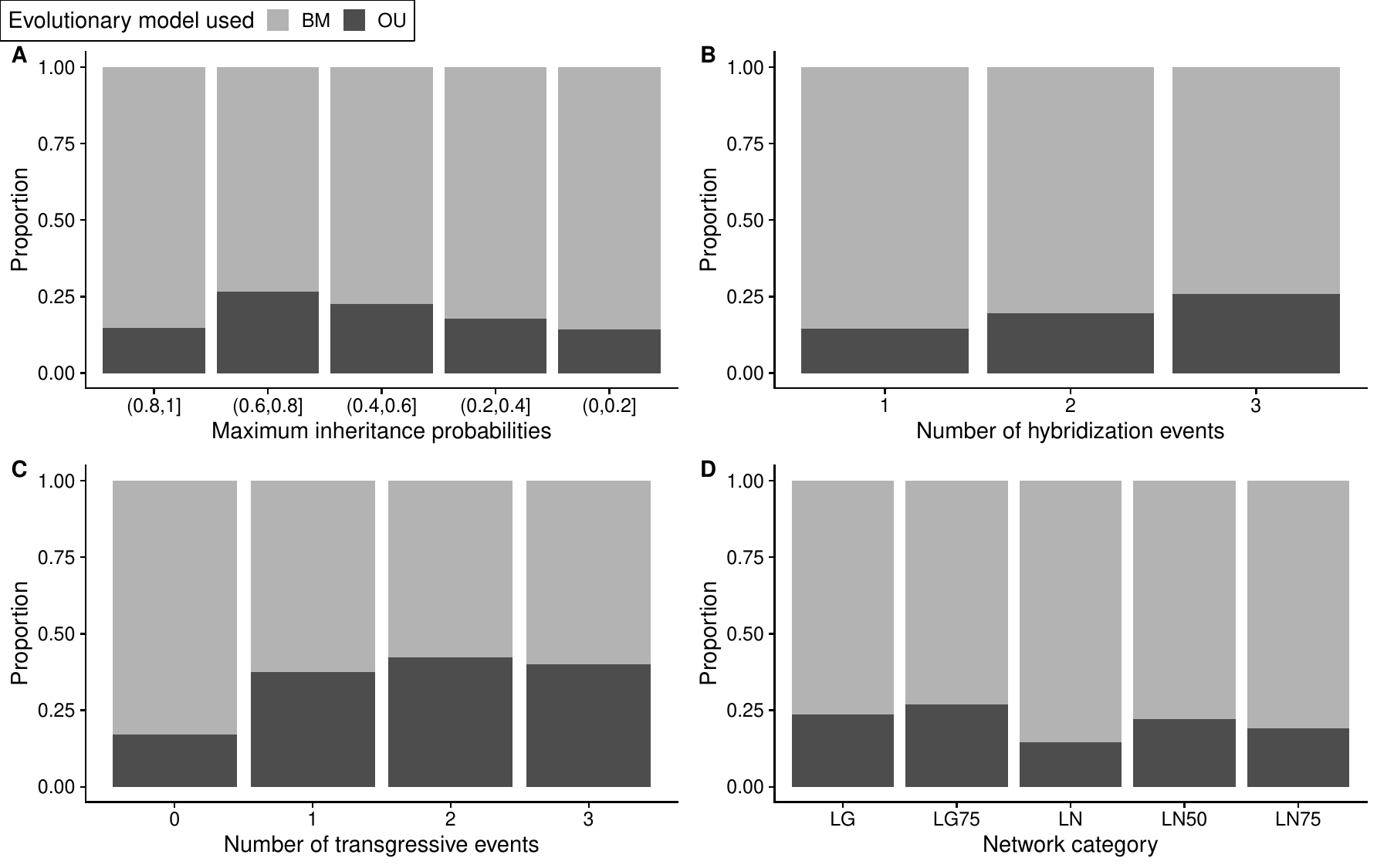}
    \caption{
    Proportion of BM (light gray) selected as models based on AIC across different simulation sets.
    (A) OU was chosen most frequently when the maximum inheritance probabilities were between 0.4 and 0.6.
    (B) The proportion of OU as the (incorrectly) selected model increased with the number of hybridization events.
    (C) OU was selected $\approx 50\%$ of the time when there was at least one transgressive event.
    (D) Networks that have lineage-neutral (LN) hybridization events show the lowest rate of selecting the OU model.
    See Table~\ref{vars_Description} for the acronyms used on the x-axis.}
    \label{fig:model-choice-vars}
\end{figure}

Figure~\ref{fig:variables-bm-vs-out} shows the results of permutation-based importance tests for the variables and differences in mean values among group-specific distributions.
Maximum inheritance probabilities (\texttt{inherMax}) contributed most to decreases in log-likelihood scores upon permutation, followed by the number of hybridization events (\texttt{nHyb}) and the number of transgressive events (\texttt{numTrans}) (Fig.~\ref{fig:variables-bm-vs-out}A). 
These results indicate that features of network structure strongly influence model performance as illustrated in Figure~\ref{fig:model-choice-vars} which shows the proportion of BM (light gray) selected as the best-fitting model based on AIC across simulation sets. Maximum inheritance probabilities differed significantly between simulations selecting OU and BM groups (Wilcoxon rank-sum test, $p < 2.2 × 10^{-16}$; Fig.~\ref{fig:model-choice-vars}A). Notably, 47\% of OU groups exhibited mean and maximum inheritance probabilities between 0.4 and 0.6. The rate of model misspecification also increased as the number of hybridization (Fig.~\ref{fig:model-choice-vars}B) and transgressive evolution events (Fig.~\ref{fig:model-choice-vars}C) increased, with OU being chosen nearly 50\% of the time at one or more transgressive evolution events. Moreover, the number of transgressive events differed significantly between OU and BM groups (Welch t-test, $p < 2.2 × 10^{-16}$), with OU groups exhibiting higher values on average. 
Networks with high LG also appeared to increase rate of misspecification, with pure LN networks exhibiting the lowest rate of model misspecification (Fig.~\ref{fig:model-choice-vars}D). 

Ornstein-Uhlenbeck model was favored when the numbers of hybridization events were increased with more symmetrical parental inheritance at reticulations. In general, these scenarios often mimic stabilizing selection by pulling trait values of seemingly independent branches towards a central optimum, differing from expectations under pure BM. 
Transgressive evolution may create a similar pattern, but for a different reason: when transgression occurs, lineages may take on phenotypes that are more similar to less closely-related lineages, leading to displacement outside of pure BM parameters and finding greater statistical support in models that explicitly include a directional pull towards some specific character state. 
Finally, LG pattern add apparently independent branches with hidden reticulate histories, thus distorting expected trait space for increased numbers of extant taxa, compared to LN pattern.

\section*{Discussion}\label{sec4}

\subsection*{Trait Estimation Error}
Overall, we found trait estimation error was only marginally higher when traits were generated under networks compared to those generated using the major tree of the networks as the model. Simultaneously, error distributions between tree- and network-based scenarios exhibited significant heteroscedasticity, indicating that researchers should expect decreases in precision accompanied by slight decreases in accuracy when using tree-based methods on traits generated under a network. 
This may mean that tree-based methods perform adequately enough for a variety of goals, especially given the benefit of a lower overall computational load. However, we found that error increases significantly when evolutionary rate is high, transgressive evolution or hybrid speciation has occurred, parental contribution is uneven, or internal branches are short. Naturally, nodes that occur near reticulation nodes and/or significant transgressive events tend to suffer from the highest estimation error. More specifically, since short internal branches means there is less time for evolutionary changes to occur, trait estimation suffers when sudden and unexpected trait shifts caused by reticulation.
Consider branch lengths: when time between speciation events is small (short internal branch lengths), there is less time for change to occur. So, when reticulation causes sudden and unexpected trait shifts, trait estimation suffers.
{Short internal branches likely also lead to a greater probability of reticulation occurring on terminal versus internal branches, where more recent trait shifts lead to increased observed displacement at tips. }
When internal branches are longer, however, shifts in trait space may be able to be accounted for by modifying other model parameters (e.g., $\sigma^2$) while maintaining some accuracy ACEs. 
When terminal branches are short and tip states are displaced by hybridization, error ensues, but perhaps with greater impact than short internal branches, as estimation error is propagated down the tree. 
Rate of trait evolution and shifts in traits due to transgressive evolution have similar effects: When under the influence of reticulation, these factors substantially shift trait values away from what would be expected under BM. In this case, our models adjust parameters and estimates of ancestral characters to maximize the likelihoods of observed tip states. 
 
In general, error follows when incorrect phylogenetic models are used to estimate trait histories \citep{Wilson2022Chronogram}. We should therefore expect error to amplify altogether when the assumed topology in ACE is not the major tree of the true evolutionary scenario represented in a network. 
This, of course, can happen even without reticulation; the history of a trait depends on the histories of the genes that underlie it, and we are well aware that much of the genome exhibits a history different from that of the population as a whole (i.e., gene tree--species tree discordance; \cite{Maddison1997GeneTrees,Szollosi2015Inference}).  
In this paper, we consider situations in which the trait history matches the population history, both are impacted by reticulation, and trait inheritance and phenotype expression follow parental inheritance proportions. 
There are likely many scenarios, however, in which trait histories don't adhere to the overarching species network, but rather follow a pattern of inheritance determined by one of the many possible sub-trees. This begs the question: Which traits should we expect to follow reticulate evolutionary patterns? The answer, like many in ecology and evolution, is likely idiosyncratic with respect to study systems and trait natures. For example, systems that exhibit large tracts of non-recombining structural variants may produce traits that do not lead to intermediate inheritance under reticulation scenarios, but instead create ``either or'' patterns, where the trait always follows one of the sub-trees. Future work should explore the extent to which traits with varying underlying historical and genomic structures are expected to exhibit tree- versus network-like inheritance patterns in the face of hybridization. Additionally, this work focused exclusively on continuous traits, so an important extension of this work will be understanding how discrete and/or binary trait estimation methods are impacted by introgression. 

\subsection*{Rate Estimation Error}
Overall,  $\sigma^2$ tends to be overestimated when trait histories follow a network compared to a tree. 
We found a variety of factors, both related to network topology and trait history, that affect the magnitude of this error. 
When models fail to account for non-independent branches, $\sigma^2$ will be inflated to account for apparently rapid observed shifts from expected values under treelike BM. Unsurprisingly, then, we found that the number of transgressive events was highly significant for GLMs, and was the most important variable under random forest analyses.
In this way, shifts in trait space produced by transgressive events can only be accounted for by increasing the rate parameter. Network-based models that allow for trait shifts at hybrid nodes outside of parental values can explicitly model these shifts without compromising accuracy and precision of parameter estimation. This, however, depends on accurate network reconstruction and hybrid node placement, otherwise even network-based methods will be prone to error \cite{bastide2018inference}. 

Interestingly, network structural variation can have both negative and positive impacts on parameter estimation: longer branches may be able to absorb some of the variation induced by hybridization, and lead to decreased overall estimation error, while shorter branches amplify the confounding effects of hybridization-induced shifts in trait space, causing increases in rate overestimation.
On the other hand, as genomic contribution of one parent increases over the other, rates are more likely to be overestimated. In particular, when parental inheritance induce stronger trait shifts toward one parent's trait-space, it can create uneven patterns of trait evolution that are best explained by increased rates under tree-based BM. Similarly, as true rates of trait evolution increase, they may potentiate shifts in trait space at hybrid nodes of increased magnitude relative to nearby tree nodes that inflate rate estimates.  

Typically, BM relies on $\sigma^2$ to account for variation in trait values given a tree. When actual evolutionary history of the trait does not match the provided phylogeny (e.g., using a tree instead of a network), $\sigma^2$ estimation becomes less reliable. Indeed, \cite{Wilson2022Chronogram} reify the notion that ancestral character estimates are most reliable when conducted using trees (with branch lengths) that are most correlated with the evolution of the character. 
Additionally, the presence and prevalence of incomplete lineage sorting can impact branch lengths and topology, and it is likely to be of greater concern when time between speciation events is smaller and ancestral population sizes are larger, impacting reliability of ancestral character and model parameter estimates \citep{LiSteelZhang2008MoreTaxa,Hahn2016,Hibbins2023Phylogenomic}. 
Because of this, our study represents the best-case scenario for reticulate topologies. It is likely that error in topological and branch length estimations arising from reticulations will exacerbate many of the problems discussed herein. Here, we show that even when the accurate major tree is recovered, reticulation still significantly impacts the reliability of ancestral character and model parameter estimates. 
Therefore, cautious interpretation of parameter estimates is necessary if the assumed phylogeny is suspected to be inaccurate as a reflection of the true evolutionary nature of focal traits. 

\subsection*{Model Choice}
Phylogenetic comparative methods have historically used the results from model-fit statistics as an indication that the chosen model is more likely to be the true model. 
We often end up in scenarios where a certain hypothesis is proposed, for example, ``trait $i$ is causally related to trait $j$'' , ``Trait $i$ causes increased diversification'', or ``Trait $i$ is evolving under selection''. 
The likelihood of sampled trait data is assessed under different models, and the model with the highest likelihood is then assumed to best represent the true evolutionary process under which the trait evolved. This is problematic, as model fit statistics are not indicative of reality, but of relative goodness of fit \textit{compared to the other models tested}. This has been thoroughly discussed in the context of State Speciation and Extinction (SSE) models, which explicitly account for changes in diversification associated with single traits. Many authors have found that some of these models (e.g., BiSSE) fail to account not only for unmodeled traits that are actually driving observed patterns \cite{Beaulieu2013}, but also fail to accommodate scenarios in which diversification is variable, but is independent of the modeled trait altogether \cite{Rabosky2015}. Essentially, model adequacy tests have found that when diversification is variable, a(n) (SSE) model that accounts for it will always outperform a model that does not, even if the trait-history relationship presumed by the model does not hold. Similarly, when a two independent traits evolve in only one clade (commonly referred to as ``unreplicated bursts''; \cite{Maddison2015PhylogeneticCorrelation}), PCMs can often return significant results with biologically meaningful implications, when the true cause of observed trait associations or diversity inequality is nothing but phylogenetic inertia. 

Our findings demonstrate a similar pattern in the case of reticulate evolution: We found that models that imply some level of stabilizing selection (i.e., OU) will outperform neutral models (i.e., BM) approximately 21\% of the time, when traits evolve under BM on networks but are modeled on bifurcating trees. The preference of OU for traits generated under reticulate evolution was significantly higher than its preference under tree-based trait evolution, indicating that branch non-independence often creates patterns that cannot be accommodated by the assumptions of tree-based BM, likely by displacing trait values in a way that mimics stabilizing selection or convergence. Our interpretation for this observation is that a neutrally evolving trait impacted by non-treelike evolution may fit better to a model that assumes stabilizing selection simply because that model accounts for observed displacements in character states that diverge from expectations under tree-based BM. Thus, we advise researchers to be wary of interpreting model fit statistics as support for biological conclusions, especially if historical reticulate evolution is possible within a focal group.

\subsection*{Modern Solutions to Age-Old Problems}
Our study has shown ample opportunity for error when true patterns of evolution violate the assumptions made by our models. Here, we specifically examine branch non-independence, or reticulation. When reticulate scenarios involve symmetrical parental inheritance ($\gamma\approx0.5$), transgressive evolution, rapidly evolving traits, or hybrid speciation rather than introgression; ancestral character estimations, evolutionary rate parameters, and models of trait evolution are likely to be misled. Further, nodes that are closest to hybridization events, and nodes on shorter branches are most prone to error. However, we also found that, in general, tree-wide trait estimation error largely reflects a decrease in precision, rather than accuracy. This is in line with the results of \cite{Bastide2018} who found that while their network-based PCM approach outperformed tree-based PCMs in estimation of ancestral characters and key model parameters, results were not so different from those generated under tree-based approaches as to invalidate previous findings. While our study expands previous work on this topic by pointing to specific features of trait and phylogenetic histories that impact variation in tree-based PCM error, it also reifies the notion that, for the most part, tree-based PCMs may lack accuracy and precision, but they don't often grossly misestimate ancestral character states when the true evolutionary history is a network. In other words, despite potentially specifying the wrong model, and estimating incorrect parameters, we can still get pretty close to the right answer. 

We should note that erroneous estimates from ACE due to violations of model assumptions are not unique to reticulate evolution.
For example, \cite{Uyeda2018} showed that when a singular evolutionary event differs drastically in rate, nature, or magnitude from other evolutionary events along the tree, PCMs (including ACE) may lose power and reliability. In general, scenarios such as singular evolutionary events, unreplicated bursts, and mismatches between trait and population histories may cause unresolvable quandaries that are difficult or impossible for current methods to resolve.
If these issues are pervasive in our data, we may often lack the statistical power to confidently infer meaningful relationships between variables (or between variables and the tree).

One possible path forward suggested by many authors including \cite{OMeara2012} and \cite{Uyeda2018} is simply to stop taking significant results from singular models or tests as indicative of a specific causal chain of events.
We should not only assess the relative fit of our models, but also the adequacy of our models. Thus, we should consider all initial results as hypotheses that can inform directed experiments, in which we test causal hypotheses against other potential explanations for the observed patterns. 
In this way, one might simulate evolutionary trajectories under different scenarios with different explanatory models, and test whether the observed results from the original data align significantly with the distribution of simulated results for competing explanatory models \citep{Matzke2022}. \cite{Uyeda2018} also suggest that researchers should be explicit about the causal processes they aim to infer, by employing methods based on structured causal frameworks such as path analysis or structural equation modeling, for example.
In this vein, while we may be adding more fuel to the fire that is PCM model evaluation, we also hope to build on this established framework by exploring the kinds of reticulation scenarios that impact conclusions from tree-based PCMs. By describing why and how phenomena related to reticulation affect our modeling results, we can create actionable and integrative plans that explicitly account for a greater diversity of evolutionary events.

So, when should we use network-based approaches instead of tree-based approaches? The answer, of course, is complicated by the fact that network construction and calibration remains computationally intensive, and many researchers simply don't have the requisite computational or genomic resources to use network-based approaches even if they suspect that reticulation has played a role in the history of their study system. Our results, however, indicate that tree-based approaches are more likely to under-perform under specific circumstances: (a) when traits evolve rapidly, (b) when there are multiple reticulation events, (c) when transgressive character displacement has occurred, (d) when time between speciation events is short, (e) when reticulation events result in hybrid speciation, and (f) when parental genetic contribution is symmetrical. If histories of introgression are unknown, we suggest beginning by using simple statistical tests, such as Patterson's D, to determine whether pursuing network-based approaches may be warranted. Thus, researchers can leverage these criteria with their own system-specific biological expertise to make informed decisions about both modeling choices and the interpretation of their results. 

In general, our attempts to model evolutionary events are likely always going to be overly simplistic and sometimes wrong. Which traits are important, and which events to model will always be a debate, and we may readily accept tidy wrong answers because they’re easier to interpret than messy right answers. We can always strive, however, to improve our methods and better understand why and how they may be misleading. While we can’t ensure that we’re going to be right, we can take steps to generate hypotheses that are less likely to be wrong. Thus, our ultimate suggestion is that we shouldn't avoid using tree-based PCMs altogether; they play an important role in unraveling the complex relationships between organisms, environments, traits, and evolutionary history.  Instead, we should proceed with caution. We should use network-based approaches when feasible, and we should be especially dubious about generating hypotheses that depend on interpreting estimated parameter values or model choice statistics as biologically meaningful when using tree-based PCMs in situations where non-treelike evolution has been observed, is highly likely, or even suspected to influence focal traits.

\section*{Acknowledgements}
All authors would like to thank the organizers and instructors of the Marine Biological Laboratory's Workshop on Molecular Evolution in 2023 at Woods Hole, MA, USA, where the seed of this paper was planted. We would also like to thank members of the Spalink lab (Texas A\&M University). Their provided expertise, novel insight, support, and guidance were invaluable to the construction of this paper. LM was supported by the Future Faculty Fellowship at Texas A\&M, and NSF-1902064. ESL was supported by the National Science and Engineering Research Council (NSERC) PGS-D scholarship. This work was also partially supported by the NSF under Grant DMS-1929284 while SK was in residence at the Institute for Computational and Experimental Research in Mathematics in Providence, RI, during the Theory, Methods, and Applications of Quantitative Phylogenomics program. This work was also partially supported by the RIKEN iTHEMS.

\section*{Data Availability Statement}
Data Availability: All the original data and scripts necessary to reproduce the analyses reported in this study can be accessed through the Dryad link: \url{http://dx.doi.org/10.5061/dryad.[NNNN]} (to be updated once accepted).
All scripts used for network, tree, and trait simulation and analyses are also available on GitHub repository: 
\url{https://github.com/lydMor/NetworksPhyloCompMethods}.
\bigskip\bigskip








\end{document}